\documentclass[conference,10pt]{IEEEtran}
\usepackage{epsfig,epsf,rotating,setspace,latexsym,amsmath,amssymb,amsfonts,bm,theorem,subfigure,epstopdf}
\usepackage{cite,authblk}
\usepackage{bbm}
\usepackage{algorithm}
\usepackage{color}
\usepackage{mathtools}
\usepackage{soul}

\newtheorem{theorem}{Theorem}

\newenvironment{Proof}[1]{\medskip\par\noindent{\bf Proof:\,}\,#1}{{\mbox{\,$\blacksquare$}\par}}

\allowdisplaybreaks

\begin{document}
 
\title{Timely Gossiping with File Slicing and \\ Network Coding}
 
\author{Priyanka Kaswan \qquad Sennur Ulukus\\
    \normalsize Department of Electrical and Computer Engineering\\
    \normalsize University of Maryland, College Park, MD 20742\\
    \normalsize  \emph{pkaswan@umd.edu} \qquad \emph{ulukus@umd.edu}}
 
\maketitle

\begin{abstract}
We consider a system consisting of a large network of $n$ users and a library of files, wherein inter-user communication is established based upon gossip mechanisms. Each file is initially present at exactly one node, which is designated as the file \emph{source}. The source gets updated with newer versions of the file according to an arbitrary distribution in real time, and the other users in the network wish to acquire the latest possible version of the file. We present a class of gossip protocols that achieve $O(1)$ age at a typical node in a single-file system and $O(n)$ age at a typical node for a given file in an $n$-file system. We show that file slicing and network coding based protocols fall under the presented class of protocols. Numerical evaluation results are presented to confirm the aforementioned bounds.
\end{abstract}

\section{Introduction}

Last couple of decades have seen a surge in different types of large communication networks, e.g., dense sensor networks, mobile ad-hoc networks, content distribution networks, autonomous driving networks, and so on, where the network unpredictability and resource-constrained infrastructure calls for employment of decentralized algorithms for efficient communication of information across the network. 

Gossip algorithms are decentralized algorithms where two nodes of the network randomly come into contact and communicate exclusively based on the limited information available at the liaising nodes. As opposed to centralized protocols, where a centralized server coordinates or assists the exchange of files between nodes of the network, in gossip algorithms, the nodes are unaware of the information available at other nodes and all actions by a node are taken solely based on its local status or information obtained from its neighbors. The idea of gossip protocols was first mentioned in \cite{Demers1987EpidemicAF} and since then there have been numerous works on gossip protocols, e.g., \cite{yaron03thesis, vocking2000, Pittel1987OnSA, deb2006AlgebraicGossip, devavrat2006, Sanghavi2007GossipFileSplit, Yates21gossip, baturalp21comm_struc, Bastopcu21gossip}.

In this respect, the works that are directly related to our work are \cite{deb2006AlgebraicGossip, devavrat2006, Sanghavi2007GossipFileSplit}. In \cite{deb2006AlgebraicGossip}, dissemination time of $k$ messages in a large network of $n$ nodes is studied for gossip protocols based on random linear coding (RLC), random message selection (RMS), and sequential dissemination. \cite{deb2006AlgebraicGossip} shows that RLC-based protocol has superior performance and has $ck + O(\sqrt{k}\log k \log n)$ dissemination time in complete graphs. \cite{devavrat2006} further extends the result to arbitrary graphs. \cite{Sanghavi2007GossipFileSplit} studies the dissemination of a file in a large network of $n$ nodes by dividing the file into $k$ pieces. Therein, the authors propose a hybrid piece selection protocol, called INTERLEAVE, which achieves a $O(k+\log n)$ total dissemination time for a file. 

However, all these works focus on using the dissemination time of a specific message or a set of messages to all nodes of the network as the performance measure. In most practical networks, information is dynamic in nature, and all nodes need to be continuously delivered with the latest file versions in real time. Therefore, in such networks, the age of information (AoI) of messages at the nodes proves to be a more relevant performance metric. AoI has been studied in a range of contexts, such as, queueing networks, energy harvesting systems, caching systems, web crawling, scheduling problems, remote estimation, UAV systems, and so on, see e.g., \cite{Najm17, Soysal19, kolobov19, Farazi18, Baknina18, Leng19, Arafa20, Elmagid20, Ceran18, liu18, elmagidUAV19, Bastopcu20_soft_updates, Bastopcu20_group, yates19_status_update, Buyukates19_hier, Bedewy19, Buyukates19_multihop, wang19_counting, bastopcu20_google, sun17_remote, Bastopcu20_infection, yun18, kam20, chakravorty20, Bastopcu20_selective, Buyukates20_stragglers, ozfatura20, yang20, Bastopcu19_distortion, Ayan19, Banerjee20, Yates17sqrt, Tang19FileAge, bastopcu2020LineNetwork, bastopcu2020LimitedCache, kaswan-isit2021, Zhang18MobileEdgeCaching, Gao12opportunistic, Pappas20EH, Eryilmaz21, Kam17HistoryFreshness, gu2020TwoHop, Arafa19TwoHop}. 

In this paper, we use \emph{version age of information} as the metric to measure the timeliness of information, which counts how many versions behind a file is at a node from the version currently prevailing at the source. A predecessor of version age was introduced in \cite{bastopcu20_google} in the context of the timeliness of the Google Scholar citation index by using the difference between the actual citation numbers and the latest updated citation numbers of the researchers on the citation index. The concept was formally defined in \cite{Yates21gossip, Eryilmaz21}.

Gossip networks have been analyzed from an information timeliness perspective in \cite{Yates21gossip, baturalp21comm_struc, Bastopcu21gossip}. \cite{Yates21gossip} shows that in a complete symmetric network of $n$ nodes with a single file, the average age at each node is $O(\log n)$. \cite{baturalp21comm_struc} demonstrates that $O(\log n)$ average version age is achievable per node in fully connected cluster models. \cite{Bastopcu21gossip} extends \cite{Yates21gossip} from the version age metric to the binary freshness metric. An interesting question to ask is whether we can do better than $O(\log n)$ version age \cite{Yates21gossip, baturalp21comm_struc} in a fully connected network. Similarly, if we disseminate $n$ files simultaneously to all $n$ nodes in real time in a network by dividing all rate resources among $n$ files in \cite{Yates21gossip}, we obtain a version age of order $O(n \log n)$. Another interesting question is whether we can do better than $O(n \log n)$ for $n$ files.

In this work, we achieve $O(1)$ version age for single-file dissemination and $O(n)$ version age for $n$-file dissemination (improving both results by $\log n$), in a network of $n$ nodes, that communicate using a random phone call model. We show that complete dissemination of a file to the entire network is not necessary to achieve these age bounds. Further, periodic pausing of dissemination of newer file versions from the source does not hinder the achievement of these bounds. We show that specific file splitting and network coding based gossip protocols achieve the mentioned bounds. Finally, we present simulation results to validate the aforementioned bounds.

\section{Gossiping with a Single File} \label{sect:single_file}

We consider slotted time, where one time slot is the duration of time needed to receive an entire file. The network consists of $n$ nodes, and a node, called the \emph{source}, receives updates for the file according to an arbitrary distribution. All source updates start at the beginning of a time slot. Hence, the source may get a new version of the file in every time slot. The $n$ users wish to have the latest possible version of the file at any time. We follow the random phone call model (i.e., rumor mongering model), where the algorithm works in rounds, and in every round, each user, uniformly at random, selects another user, called the \emph{target} node, to communicate with.

\begin{figure*}[t]
\centerline{\includegraphics[width=0.95\linewidth]{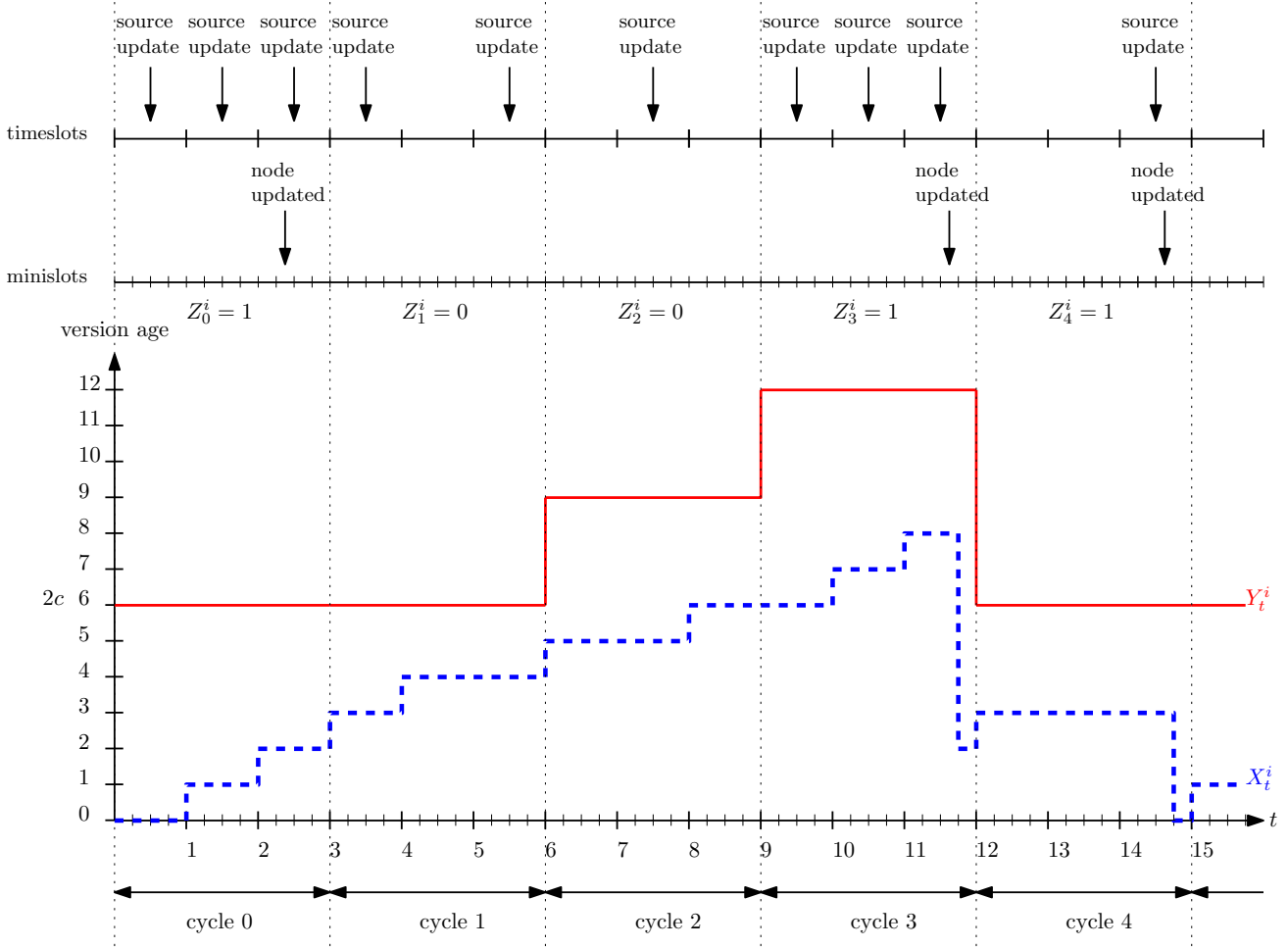}}
\caption{Sample version age evolution at typical node $i$, with parameters $c=3$, $\ell=4$. The red curve corresponds to $Y^i_t$ which remains above $2c=6$, and increases by $c=3$ whenever the node is unsuccessful in receiving the update in the preceding cycle. The blue curve corresponds to $X^i_t$ which drops to lower values in the minislot after it gathers all pieces of a particular file version.}
\label{fig:filesplitting}
\vspace*{-0.2cm}
\end{figure*}

\begin{theorem}\label{thm1}
    In a network of $n$ users, where users communicate according to a gossip protocol, if there exists $p\in [0,1)$ and $n_0 \in \mathbb{N}$, such that for all $n>n_0$, the probability of a random node receiving the file in $O(1)$ time is greater than $1-p$, then the protocol achieves an $O(1)$ age at each node.
\end{theorem}

\begin{Proof}
Let $c=O(1)$ time slots, $c \in \mathbb{N}$, which is also taken to be the duration of a cycle. The timeline is thus divided into consecutive cycles, with each cycle consisting of $c$ time slots; see Fig.~\ref{fig:filesplitting}. At the beginning of each cycle, the source begins to transmit the latest version it has received by the end of the previous time slot to the network. The network is allowed to gossip for $c$ time slots according to the given protocol. To avoid mixing of different versions of a file, when the cycle ends, the nodes lay off transmission of the file versions already in their possession, and the circulation of a newer version is set in motion with the beginning of the next cycle.

In time slot $t$, let $N^i_t$ be the latest version of the file available at node $i$, and let $N^s_t$ be the latest version present at the source. Thus, the instantaneous version age of information at node $i$ at time $t$ is $X^i_t=N^s_t-N^i_t$. Therefore, every time the source receives an update, $N^s_t$, and hence, the instantaneous age at each node in the network, increment by one.

We assume that all nodes start with zero age at $t=0$. At $t=kc$, $k\in \mathbb{N}$, the cycle $k$ begins, and the source node starts the transmission of version $N^s_{kc}$ to the network. During cycle $k$, i.e., between time slot $kc$ and $(k+1)c$, the source may receive a maximum of $c$ updates in the $c$ time slots, but it continues to circulate the version $N^s_{kc}$ and does not transmit the newer versions. If node $i$ receives the version $N^s_{kc}$ at any time $t$, $kc\leq t <(k+1)c$, then the instantaneous version age at node $i$ drops to $X^i_t=N^s_t-N^s_{kc}$. 

Let $Z^i_k$ indicate if node $i$ successfully receives the file in cycle $k$. Hence,
\begin{align}
Z^i_k = \begin{cases} 
1, & \text{if node $i$ receives version $N^s_{kc}$ in cycle $k$}\\
0, & \text{otherwise}
\end{cases}
\end{align}
Note that $Z^i_k$ subscript $k$ denotes the cycle index whereas $N^s_t$, $N^i_t$ and $X^i_t$ subscripts $t$ denote the time slot index (see Fig.~\ref{fig:filesplitting}). 

At the end of cycle $k$,
\begin{align}
X^i_{(k+1)c} = \begin{cases} 
X^i_{kc}+N^s_{(k+1)c}-N^s_{kc}, & \text{if $Z^i_k=0$}\\
N^s_{(k+1)c}-N^s_{kc}, & \text{if $Z^i_k=1$}\\
\end{cases}\label{eq:new_slot_X}
\end{align}
where we note that the source would have moved from version $N^s_{kc}$ to version $N^s_{(k+1)c}$ through cycle $k$, and $X^i_{kc}=N^s_{kc}-N^i_{kc}$. Here, observe that if $Z^i_k=0$, then $N^i_{(k+1)c}=N^i_{kc}$, and if $Z_i^k=1$, then $N^i_{(k+1)c}=N^s_{kc}$.

Next, we define $Y^i_t$ for node $i$ (see Fig.~\ref{fig:filesplitting}), where the subscript $t$ is the time slot index. For $kc\leq t <(k+1)c$,
\begin{align} 
    Y^i_t=2c Z^i_{k-1}+(1-Z^i_{k-1})(Y^i_{kc-1}+c)\label{eq:defn_Yi_t}
\end{align}
with $Y^i_t=2c$ for cycle $0$. Note that $Y^i_t$ remains constant over the duration of any cycle.

Next, we show that $Y^i_t$ forms an upper bound on $X^i_t$ for all $t$. First, note that, since the source can get a maximum of $c$ new updates in a cycle of $c$ slots, we have $N^s_{kc}-N^s_{(k-1)c} \leq c$. Then, if $Z^i_{k-1}=1$, i.e., if node $i$ is successfully updated in cycle $k-1$, from (\ref{eq:new_slot_X}), we have $X^i_{kc}\leq c$, i.e., at the beginning of the $k$th cycle, the $i$th node's version age is upper bounded by $c$. By repeating this argument, $N^s_{(k+1)c}-N^s_{kc} \leq c$, and re-employing (\ref{eq:new_slot_X}) gives $X^i_{t}\leq 2c$ throughout the $k$th cycle, irrespective of the value of $Z^i_k$, i.e., whether the $i$th node is updated or not in the $k$th cycle. Note from (\ref{eq:defn_Yi_t}) that when $Z^i_{k-1}=1$ we have $Y^i_t=2c$ for the entire $k$th cycle. Thus, when $Z^i_{k-1}=1$, we have $X^i_t \leq Y^i_t=2c$ throughout the $k$th cycle. On the other hand, if $Z^i_{k-1}=0$, since the source can get at most $c$ updates over $c$ slots, $X^i_t\leq X^i_{kc-1}+c$ for $t\in [kc,(k+1)c)$. In this case, from (\ref{eq:defn_Yi_t}), $Y^i_{t}= Y^i_{kc-1}+c$. All of these imply that $X^i_t \leq Y^i_t$ for all $t$, as graphically shown in Fig.~\ref{fig:filesplitting}. 

Next, we bound $\mathbb{E}[Y^i_t]$ which in turns bounds $\mathbb{E}[X^i_t]$ due to $X^i_t \leq Y^i_t$. Consider $kc\leq t <(k+1)c$. Recall that if node $i$ successfully receives a file in cycle $(k-1)$, i.e., $Z^i_{k-1}=1$, then $Y^i_t$ drops to $2c$, otherwise it increases by $c$. Hence, for $m \in \{2,\ldots,k+1\}$, $Y^i_t=mc$ is possible only if $Z^i_{k-m+1}=1$ and $Z^i_{j}=0$ for all $j \in \{k-m+2,\ldots, k-1\}$, $j\in \mathbb{N}$. That is, the node was updated $m-1$ slots ago, and could not get updated after that, accumulating an increment of $c$ in $Y^i_t$ value in each of the later slots. Further, if $Y^i_t=(k+2)c$, then that means that node $i$ failed to receive any updates since $t=0$. Since $Z^i_k$ are i.i.d.~in $k$ for node $i$, define $p_n=\mathbb{P}(Z^i_k=0)$. Since we consider a symmetric fully connected network, all nodes have the same per cycle failure probability $p_n$ in an $n$ node network. Hence, we have the following bound,
\begin{align}
    \mathbb{E}[Y^i_t]&=\sum_{m=2}^{k+1} mc(1-p_n)p_n^{m-2}+(k+2)cp_n^{k}\\
    &<\frac{c}{p_n}\sum_{m=2}^{k+2}mp_n^{m-1}\\
    &<\frac{c(2-p_n)}{(1-p_n)^2}\\
    &=\frac{c}{(1-p_n)^2}+\frac{c}{(1-p_n)} \label{laststep}
\end{align}

Note that the expression in (\ref{laststep}) is an increasing function of $p_n$. As the statement of the theorem states $\mathbb{P}(Z^i_j=1)>1-p$ for all $n>n_0$, thus $p_n<p$, and consequently 
\begin{align}
    \mathbb{E}[Y^i_t]<\frac{c}{(1-p)^2}+\frac{c}{(1-p)}
\end{align}
Hence, $\mathbb{E}[X^i_t]<\mathbb{E}[Y^i_t]<\frac{c}{(1-p)^2}+\frac{c}{(1-p)}$ for network size $n>n_0$, and the protocol achieves $O(1)$ age at each node. 
\end{Proof}

We have made several relaxations and sub-optimalities in the analysis above. Before we look back at all these relaxations, we first consider an example of such a gossip protocol, the INTERLEAVE protocol proposed in \cite{Sanghavi2007GossipFileSplit}. Motivated by file splitting employed in protocols such BitTorrent, INTERLEAVE is a file splitting based dissemination scheme that is able to satisfy the conditions of Theorem~\ref{thm1} as we show next. 

\subsection{An Achievable Gossiping Protocol: INTERLEAVE}

The INTERLEAVE protocol of \cite{Sanghavi2007GossipFileSplit} employs an alternating combination of push and pull actions at each user in the network to disseminate a file. The file is split into $\ell$ pieces by the source, and the pieces are numbered before dissemination to the network. Each piece consumes $\frac{1}{\ell}$ fraction of a time slot for transmission, which we define to be one \emph{minislot}. Thus, in the timeline, each time slot is further divided into $\ell$ minislots. 

In an odd minislot, each user chooses a target node uniformly at random from the network, and pushes to its target node the highest numbered piece it has received in the previous odd minislots. The source pushes a fresh piece to a target node in the network in every odd minislot, where the pieces are chosen  in the order they are numbered. In an even minislot, each user again chooses a target node uniformly at random from the network, and makes a pull request for the lowest numbered piece that is missing in its collection from its target node. Node $i$ is said to have received a particular version of the file when it has accumulated all pieces of that version.

\cite[Thm.~6]{Sanghavi2007GossipFileSplit} provides that INTERLEAVE disseminates all $\ell$ pieces of a file to all $n$ nodes of the network within $9(\ell+\log n)$ minislots, with probability $1-5n^{-s}$, for any $s<0.5$. If we choose $\ell \geq \alpha \log n$ for some constant $\alpha>0$, then considering a minislot to be $\frac{1}{\ell}$ duration of a full time slot, INTERLEAVE takes $9(1+\frac{\log n}{\ell})=O(1)$ time for spreading a particular file to the entire network with high probability. Since the probability $1-5n^{-s}$ is an increasing function of $n$, the INTERLEAVE protocol satisfies the requirements of Theorem~\ref{thm1} and hence can help achieve $O(1)$ expected age at any node in the network.

The inner-workings of INTERLEAVE may look complicated at first, and it may look as though it requires more information than the gossip algorithm in \cite{Yates21gossip}. To this, we point out that, in INTERLEAVE all nodes are oblivious of the collection of file pieces existing at other nodes of the network. Each node makes a pull request for the lowest numbered piece missing in its private collection, and pushes out the highest numbered piece in its private collection. Therefore, the decision making in INTERLEAVE is independent of the information at other nodes, and thus, the INTERLEAVE protocol is a gossip protocol, as the one in \cite{Yates21gossip}.

Next, we note the relaxations made in the calculations in Theorem~\ref{thm1}: 1) The probability of successfully receiving a file in a cycle, $\mathbb{P}(Z^i_k=1)$, is replaced by its lower bound $1-p$. 2) At the beginning of each new cycle, all nodes are made to halt transmissions of all files in their possession uptill the previous cycle. This could result in some nodes sitting idle in the beginning of a cycle waiting for a piece of the newer version, while they could have continued disseminating pieces of previous version missing in other nodes. 3) $Y^i_t$ is a crude upper bound to $X^i_t$; $Y^i_t$ only changes value at the end of a cycle, though it is possible for node to gather all pieces of a file before the end of the cycle. 4) The evolution of $Y_t^i$ assumes the worst case scenario where the source gets the maximum possible $c$ version updates in every cycle, and hence file version received by the end of the cycle is always $c$ slots old. Further, $Y_t^i$ is set constant over cycle $k$, upper bounding the maximum value of $X_t^i$ within that cycle. 5) The protocol may take less than $c$ time slots for larger values of $n$ for disseminating the file to all nodes with same probability. Hence, use of $c$ time slots in a cycle might be sub-optimal.

Finally, as a general remark on Theorem~\ref{thm1}, interestingly, a protocol can achieve $O(1)$ age even if the success probability $1-p$ has a very small value, causing very fewer nodes to really receive a file in a particular cycle. One reason for this can be that every cycle presents a fresh opportunity for a node to receive a new version and reduce its age. Hence it is okay to miss out some updates in a couple of cycles.

\section{Gossiping with Multiple Files} \label{sect:multiple_file}

We consider a system consisting of a network of $n$ nodes and a library of $n$ files. Each node is responsible for creating version updates for a unique file. We label the node acting as source for file $f$ as node $f$. Then, file $f$ is termed as the source file for node $f$. We consider slotted time, and define one time slot to be the time required to receive an entire file. All file transmissions start at the beginning of a time slot and consume an entire time slot. We follow the random phone call model, where in every time slot, each user uniformly at random picks another user, referred to as the target, and communicates with either a pull or a push action based on a gossip protocol. 

Node $f$ receives version updates for its source file $f$ according to an arbitrary distribution. Thus, in each time slot, node $f$ may or may not get a newer version update. The updates at all nodes for their respective source files are considered uncoordinated, implying that some files may get updated more frequently at their respective source nodes than others. All users wish to obtain the latest possible version of all files at any point of time. 

\begin{theorem}\label{thm2}
    In a network of $n$ users and $n$ files, where users communicate according to a gossip protocol, if there exists $p\in [0,1)$ and $n_0 \in \mathbb{N}$, such that for all $n>n_0$, the probability of a random node receiving all $n$ files in $O(n)$ time is greater than $1-p$, then the protocol achieves $O(n)$ age for each file at each node.
\end{theorem}

\begin{Proof}
Let $cn=O(n)$ time slots, $c \in \mathbb{N}$, which is also taken to be the duration of one cycle. The timeline is divided into consecutive cycles where each cycle consists of $cn$ time slots. At the beginning of each cycle, all nodes begin dissemination of the latest version of their respective source files received by the end of the previous time slot to the network. The network is allowed to gossip over one cycle or $cn$ time slots according to the given protocol. To avoid mixing of multiple versions of different files, at the end of the cycle, all nodes lay off transmission of file versions already in their possession, and circulation of newer versions of all files starts with the beginning of the next cycle.

In time slot $t$, let $N^{if}_t$ denote the latest version of file $f$ available at node $i$ and let $N^f_t$ be the latest version of file $f$ at its source node $f$. The instantaneous version age for file $f$ at node $i$ is $X^{if}_t=N^f_t-N^{if}_t$. Therefore, every time source node $f$ is updated with a newer version for file $f$, the instantaneous version age at every other node for file $f$ increments by one. 

The evolution of $X^{if}_t$ is similar to Section~\ref{sect:single_file}. At the beginning of cycle $k$, each node $f$ begins transmission of version $N^f_{knc}$ of its source file $f$ to the network. During the cycle, every source node can receive a maximum of $cn$ updates for its source file. The source nodes hold back the newer versions and so the system continues circulation of version $N^f_{knc}$ for each file $f$. We define $Z_k^{if}$ for cycle $k$ as follows
\begin{align}
Z^{if}_k = \begin{cases} 
1, & \text{node $i$ receives version $N_{knc}^{f}$ in cycle $k$}\\
0, & \text{node $i$ does not receive $N_{knc}^{f}$ in cycle $k$}
\end{cases}
\end{align}

For upper bounding the expected version age at time $t$ for file $f$ at node $i$, $\mathbb{E}[X^{if}_t]$, we define $Y_t^{if}$, for $kcn\leq t <(k+1)cn$, as follows
\begin{align}
    Y^{if}_t=2ncZ^{if}_{k-1}+(1-Z^{if}_{k-1})(Y^{if}_{knc-1}+nc)\label{eq:defn_Yif_t}
\end{align}
with $Y^{if}_t=2nc$ for cycle $0$. Note that $Y^{if}_t$ remains constant over the duration of any cycle. 

Since any source node can get a maximum of $nc$ new updates in a cycle of $nc$ slots for its source file, $N^f_{(k+1)nc}-N^f_{knc} \leq nc$. Using arguments similar in the proof of Theorem~\ref{thm1} allows us to conclude $Y^{if}_t \geq X^{if}_t$ for all $t$. In turn, $\mathbb{E}[Y^{if}_t]$ can be upper bounded as in Theorem~\ref{thm1}, as given below
\begin{align}
    \mathbb{E}[Y^{if}_t]<\frac{cn}{(1-p)^2}+\frac{cn}{(1-p)}
\end{align}
Hence $\mathbb{E}[X^{if}_t]<\frac{cn}{(1-p)^2}+\frac{cn}{(1-p)}=O(n)$.
\end{Proof}

\subsection{An Achievable Gossiping Protocol: RLC}
We now show that the random linear coding (RLC) gossip protocols proposed in \cite{deb2006AlgebraicGossip} fall under the category of such protocols, which work on the principle of algebraic mixing of files for faster simultaneous dissemination of all files. In RLC with PUSH protocol, each node creates a linear combination of all the coded packets in its possession at the beginning of each time slot, and transmits it to a randomly chosen target node. The coefficients are randomly chosen from the finite field $\mathbb{F}_q$, with $q \geq n$. In RLC with PULL protocol, nodes pull such random coded messages from their target nodes in every time slot. In both protocols, once a node has collected $n$ independent linear combinations, it can decode all messages.

\cite[Thm.~3.1 and Thm.~3.2]{deb2006AlgebraicGossip} provide that both RLC protocols can disseminate $n$ files to $n$ nodes in $O(n)$ time with high probability $1-O(\frac{1}{n})$. Since the probability $1-O(\frac{1}{n})$ is an increasing function of $n$, both RLC protocols satisfy conditions of Theorem~\ref{thm2} in this paper.

\section{Numerical Results}

\begin{figure}[t]
\centerline{\includegraphics[width=0.9\linewidth]{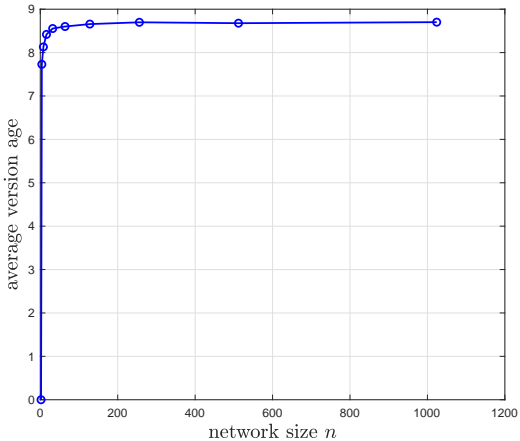}}
\vspace*{-0.3cm}
\caption{Version age versus network size $n$ with INTERLEAVE protocol.}
\label{fig:filesplitting_age}
\vspace*{-0.4cm}
\end{figure}

We now further validate our results with numerical evaluations. First, we simulate single-file dissemination using the INTERLEAVE protocol in real-time over $200$ cycles. We choose number of pieces $\ell=\lfloor \log_2 n \rfloor$ and $c=18$ in Section~\ref{sect:single_file}. Since $\ell$ changes value only when $n$ crosses a power of two, we choose $n=2^m$, $m\in \mathbb{N}$ to have $\ell$ increase consistently with $n$. We assume the source node receives version updates according to Bernoulli(0.7) distribution in every time slot. Fig.~\ref{fig:filesplitting_age} shows the expected version age per node as a function of network size $n$. We observe that as the network size grows larger, the version age does not increase, and remains below $9$ for the choice of parameters above. For $c=18$, the upper bound derived in Theorem~\ref{thm1} is $\frac{c}{(1-p)^2}+\frac{c}{(1-p)} \geq 36$, which shows that the derived upper bound is an overestimate.

\begin{figure}[t]
\centerline{\includegraphics[width=0.9\linewidth]{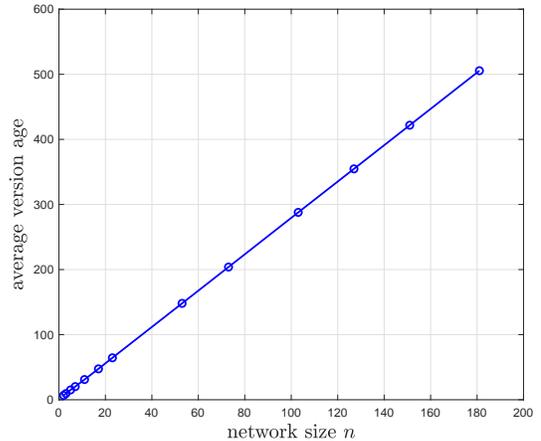}}
\vspace*{-0.3cm}
\caption{Version age versus network size $n$ with RLC with PUSH mechanism.}
\label{fig:networkcoding_age}
\vspace*{-0.4cm}
\end{figure}

\begin{figure}[t]
\centerline{\includegraphics[width=0.9\linewidth]{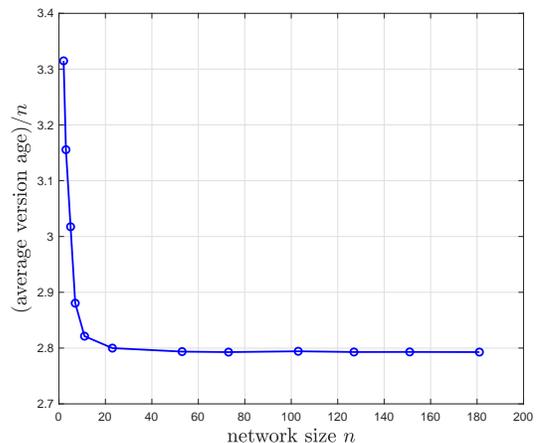}}
\vspace*{-0.3cm}
\caption{(Version age)/$n$ versus $n$ with RLC with PUSH mechanism.}
\label{fig:networkcoding_age_by_n}
\vspace*{-0.4cm}
\end{figure}

Next, we use RLC with PUSH mechanism for real-time dissemination of $n$ files in an $n$ node network. The authors in \cite{deb2006AlgebraicGossip} point out that their upper bound in \cite[Thm.~3.2]{deb2006AlgebraicGossip} is not tight and their simulations suggest that the dissemination time is close to $1.5n+\log_2 n$. We choose $c=6$ and run the scheme of Section~\ref{sect:multiple_file} over $100$ cycles. We assume all nodes receive version updates for their respective source files according to Bernoulli(0.7) distribution in every time slot, and choose $n$ to be prime numbers to keep field size $q=n$ as in \cite{deb2006AlgebraicGossip}. Fig.~\ref{fig:networkcoding_age} shows that per node per file expected version age grows almost linearly with network size $n$. Fig.~\ref{fig:networkcoding_age_by_n} plots version age divided by network size $n$, and we see the graph to be lower than 3 for above choice of parameters, which suggests that the upper bound in Theorem~\ref{thm2} (which is $\geq 12$) is an overestimate. We note that, in $n$-file simulations, we consider all files to be successfully decoded by a node at the same time, when $n$ linearly independent coded packets arrive at the node. For some $r<n$, the node might receive $r$ linearly independent combinations of $r$ files and decode them before receiving $n$ linearly independent combinations. Hence, Fig.~\ref{fig:networkcoding_age} is an upper bound to the actual average version age.

\section{Conclusion}
We studied classes of gossip protocols which achieve $O(1)$ age at each node in a single-file $n$-node system, and $O(n)$ age for each file at each node in an $n$-file $n$-node system. We saw that the bounds hold even though files might not be successfully received by the nodes in every cycle, as each cycle presents a new opportunity to the nodes to get updated. We demonstrated that gossip protocols based on file splitting and network coding fall under the above category of protocols. 

\newpage 
\bibliographystyle{unsrt}
\bibliography{ref_priyanka}
\end{document}